\documentclass[twocolumn,showpacs,amsmath,amssymb,amsfonts,floatfix,prl,aps]{revtex4}
\usepackage{graphicx}
\usepackage{color}

\begin{document}
\title{Spectral density and metal-insulator phase transition in Mott insulators within RDMFT}
\author{S. Sharma$^1$}
\email{sharma@mpi-halle.mpg.de}
\author{J. K. Dewhurst$^1$}
\author{S. Shallcross$^2$}
\author{E. K. U. Gross$^1$}
\affiliation{1 Max-Planck-Institut f\"ur Mikrostrukturphysik, Weinberg 2, 
D-06120 Halle, Germany.}
\affiliation{2 Lehrstuhl f\"ur Theoretische Festk\"orperphysik,
Staudstr. 7-B2, 91058 Erlangen, Germany.}
\date{\today}

\begin{abstract}
We present a method for calculating the spectrum of periodic solids within
reduced density matrix functional theory. This method is validated by 
a detailed comparison of the angular momentum projected spectral density
with that of well established many-body techniques, in all cases finding an excellent agreement.
The physics behind the pressure induced insulator-metal phase transition in MnO is investigated.
The driving mechanism of this transition is identified as increased
crystal field splitting with pressure, resulting in a charge redistribution between
the Mn $e_g$ and $t_2g$ symmetry projected states. 
\end{abstract}

\maketitle


Transition metal oxides (TMOs), the prototypical Mott insulators, are test-bed systems for new 
functionals within density functional theory (DFT) and many-body theories alike. 
Ground state spectra obtained from many-body theories are in good agreement with experiments.
Moreover, the spectral density obtained using dynamical mean
field theory (DMFT)\cite{kunes08} and the $G_0W_0$ corrected 
DFT\cite{rodl} agree with each other even for subtle features such as symmetry and site projected 
spectral density. Single particle DFT spectra can also be made to agree with these many-body 
results by using two separate fitting parameters; the on-site Coulomb term $U$ and the scissors
shift $\Delta$, where $\Delta$ is the difference between the experimental gap and the Kohn-Sham gap obtained
using the LSDA+$U$ functional\cite{rodl}.

Away from the ground-state TMOs show the rich physics of insulator-metal phase transitions. 
The classic Mott insulator, MnO, exhibits metalization under pressure. This phase transition is
accompanied by a simultaneous moment and volume collapse\cite{expt1,expt2,expt3,expt4,expt5,expt6}. On the theory side, however,
the physics of this phase transition is totally different for different methods; while DFT
results indicate that the increase in band width controls the phase transition\cite{cohen,luke}
DMFT results, on the other hand,
show that the main reason for metalization lies in the increased crystal field splitting\cite{kunes08}.

Recently, RDMFT has shown potential for correctly treating Mott insulators under ambient conditions\cite{sharma08,nek10}. 
RDMFT is an appealing alternative since it does not require any system dependent parameters 
and thus is a truly \emph{ab-initio} theory for treating strong correlations. However, it 
still remains to be seen how RDMFT performs away from ambient pressure conditions; can RDMFT
capture the insulator-metal phase transition? What is the physics of this phase transition within RDMFT? In order to
answer these questions one requires two things: (1)a magnetic extension of RDMFT and (2) information about photo-emission
spectrum to shed light on the nature of the phase transition. The
latter is a difficult quantity to extract from RDMFT which, by its very nature, is a ground-state theory.

In the present work we extend RDMFT to describe magnetic solids and further present a technique for calculating the photo-emission spectrum. 
We validate this technique by demonstrating an excellent agreement of the $t_{2g}$ and $e_g$ resolved spectral 
density thus obtained, with the well established many-body methods like $GW$ and DMFT.
We further show that not only at  ambient pressure but also away from it RDMFT correctly determines the spectra of Mott 
insulators and captures the physics of the insulator to metal phase transition.


Within RDMFT, the one-body reduced density 
matrix (1-RDM) is the basic variable \cite{lodwin,gilbert} 
\begin{align}\label{1rdm}
 \gamma({\bf r}, {\bf r'})\equiv N\int d^3r_2\ldots d^3r_N
 \Psi({\bf r},{\bf r}_2 \ldots {\bf r}_N)
 \Psi^*({\bf r}',{\bf r}_2 \ldots {\bf r}_N),
\end{align}
where $\Psi$ denotes the many-body wavefunction and $N$ is the total number of
electrons. Diagonalization of $\gamma$ produces a set of orthonormal Bloch functions, 
the so called natural orbitals\cite{lodwin}, 
$\phi_{i{\bf k}}$, and occupation numbers, $n_{i{\bf k}}$. 
In the present work we have extended RDMFT
 to the truly non-collinear magnetic case by treating the natural orbitals as two component Pauli-spinors,
leading to the spectral representation:
$ \gamma({\bf r},{\bf r}')=\sum_{i{\bf k}} 
 n_{i{\bf k}}\varphi_{i{\bf k}}({\bf r})\otimes\varphi_{i{\bf k}}^*({\bf r}')$,
with the necessary and sufficient conditions for ensemble  $N$-representability
of $\gamma$ \cite{coleman} requiring $0\le n_{i{\bf k}}\le 1$ for all $i$ and ${\bf k}$, 
and $\sum_{i{\bf k}}n_{i{\bf k}}=N$.

In terms of $\gamma$, the total ground-state energy \cite{gilbert} of the 
interacting system is (atomic units are used throughout)
\begin{align} \label{etot} \nonumber
E[\gamma]=&-\frac{1}{2} {\rm tr}_{\sigma}\int\lim_{{\bf r}\rightarrow{\bf r}'}
\nabla_{\bf r}^2 \gamma({\bf r},{\bf r}')\,d^3r'
+\int\rho({\bf r}) V_{\rm ext}({\bf r})\,d^3r \\
&+\frac{1}{2}  \int 
\frac{\rho({\bf r})\rho({\bf r}')}
{|{\bf r}-{\bf r}'|}\,d^3r\,d^3r'+E_{\rm xc}[\gamma],
\end{align}
where $\rho({\bf r})={\rm tr}_{\sigma}\gamma({\bf r},{\bf r})$, $V_{\rm ext}$ is a given
external potential, and $E_{\rm xc}$ we call the exchange-correlation (xc)
energy functional. In principle, Gilbert's \cite{gilbert} generalization of the
Hohenberg-Kohn theorem to the 1-RDM guarantees the existence of a functional
$E[\gamma]$  whose minimum, for fixed a $V_{\rm ext}$ yields the exact $\gamma$ 
and the exact ground-state energy. In practice, however, the correlation energy 
is an unknown functional of $\gamma$ and needs to be approximated. While there 
are several known approximations
for the xc energy functional, the most promising for extended systems is the
power functional\cite{sharma08} where the xc energy is given by
$E_{\rm xc}[\gamma]= -\frac{1}{2}\int \, \int d^3r' d^3r
 \frac{|\gamma^{\alpha}({\bf r},{\bf r}')|^2}{|{\bf r}-{\bf r}'|}$
with $\alpha$ indicating the power in the operator sense. In view of the universality of the functional $E_{\rm xc}[\gamma]$,
the value of $\alpha$ should, in principle, be system-independent. A few "optimum values" of $\alpha$ have been suggested in the 
literature\cite{power_finite,sharma08,esa11}.
In the present work $\alpha$ is fixed to 0.656 for all materials studied.

In order to devise a theoretical method to approximately obtain
the spectral density we start from the definition of the retarted Green's function written in the basis of 
the natural orbitals
\begin{align}\label{GF}
iG^{\rm R}_{\lambda \lambda'}(t-t')=\Theta(t-t')
\langle \Psi_0^N |\{a_{\lambda}(t), a^{\dagger}_{\lambda'}(t')\} | \Psi_0^{N} \rangle,
\end{align}
where $\lambda \equiv \{i,{\bf k}\}$ with the index $i$ labeling the natural orbitals for
a given {\bf k}, $a$, $a^{\dagger}$ are the creation and annihilation operators associated 
with the complete set of natural orbitals and $| \Psi_0^{N} \rangle$ is the neutral $N$-electron 
ground-state. The spectral function $A_{\lambda \lambda'}(\omega)$
can be written in terms of the Lehmann representation as:
\begin{eqnarray}\label{ALL}
&&A_{\lambda \lambda'}(\omega)=-2\Im G^{\rm R}_{\lambda \lambda'}(\omega) =2\pi\sum_j \langle \Psi_0^N |  a_{\lambda}| \Psi_j^{N+1} \rangle \\ \nonumber
&&\langle \Psi_j^{N+1} |  a^{\dagger}_{\lambda'}| \Psi_0^{N} \rangle \delta(\omega-[E_j^{N+1}-E_0^{N}]) +2\pi \sum_i\\ \nonumber
&& \langle \Psi_0^N | a^{\dagger}_{\lambda'} | \Psi_i^{N-1} \rangle 
\langle \Psi_i^{N-1} | a_{\lambda} | \Psi_0^{N} \rangle \delta(\omega-[E_0^{N}-E_i^{N-1}]),\nonumber
\end{eqnarray}
where $H| \Psi_i^{N\pm 1} \rangle = E_i^{N\pm 1}| \Psi_i^{N\pm 1} \rangle $ is satisfied by the exact $(N\pm 1)$-particle 
eigen-states of the Hamiltonian $H$. To deduce an approximate expression for the spectral function,
we replace the complete set of eigen-functions $\{|\Psi_i^{N\pm 1}\rangle\}$ by the set of approximate eigen-functions
obtained by adding/removing a single electron in a natural orbital to/from the exact correlated $N$-particle ground-state:
\begin{align}\label{Np1}
|\Phi_{\zeta}^{N+1} \rangle=\frac{1}{\sqrt{1-n_{\zeta}}}a^{\dagger}_{\zeta} | \Psi_0^N \rangle, \, \, \, \,
|\Phi_{\zeta}^{N-1} \rangle=\frac{1}{\sqrt{n_{\nu}}}a_{\zeta} | \Psi_0^N \rangle.
\end{align}
While these many-body states are clearly not a complete set, we do expect them to capture the dominant 
contributions to direct and inverse photo-emission. The set $\{|\Phi_i^{N\pm 1}\rangle\}$ does not include 
$(N\pm 1)$-states where in addition to adding/removing an electron, other electrons are excited from the ground-state,
i.e. terms involving more than one creation/annihilation operator. Thus we do not expect to be able to reproduce 
plasmon satellites. 
Replacing in Eq. (\ref{ALL}) the complete set of exact eigen-functions by this incomplete set of approximate eigen-functions
(in Eq. (\ref{Np1})) and using the fact that the natural orbitals diagonalize $\gamma$, i.e. 
$\langle \Psi_0^{N} | a^{\dagger}_{\lambda'}a_{\zeta}| \Psi_0^{N} \rangle =\delta_{\lambda \zeta} n_{\zeta}$, we end 
up with the following approximation for the spectral function:
\begin{align}\label{spect}
A_{\lambda\lambda'}(\omega) = 2\pi \delta_{\lambda\lambda'}\left[ n_{\lambda} \delta(\omega-\epsilon^-_{\lambda})+
(1-n_{\lambda}) \delta(\omega+\epsilon^+_{\lambda}) \right],
\end{align}
with $\epsilon^{\pm}_{\lambda}=E_0^N - E_{\lambda}^{N\pm1}$. We note that inspite of being approximate, the spectral
function in Eq. (\ref{spect}) satisfies the exact sum rule,
$\frac{1}{2\pi}\int_{\infty}^{\infty} A_{\lambda\lambda'}(\omega) d\omega = 1$.
Being, as function of $\omega$, a single $\delta$ functon for each fixed $\lambda=(i,{\bf k})$, the spectral function
in Eq. (\ref{spect}) is reminiscent of a non-interacting mean-field type approximation. We emphasize that our approximation
is by no means mean-field because the $({N\pm1})$-states in Eq. (\ref{Np1}) are correlated and may even be strongly correlated
if, for example, $|\Psi_0^{N}\rangle$ represents the ground-state of a Mott insulator.

It is a formidable task to determine quasi-particle life-times within the Lehmann representation: one needs to determine 
the position of an infinite number of peaks corresponding to an infinite number of $N\pm 1$ eigen-states in Eq. (\ref{ALL}).
This coalescence of peaks is described by an envelope function the width of which is proportional to the inverse quasi-particle life-time. 
In order to determine these life-times within the present formalism, one may use the more general form of the $N\pm 1$ 
states known from the extended Koopmans' theorem \cite{ekt},
$| \chi_j^{N\pm 1} \rangle = \sum_{\lambda} \alpha^{\pm}_{j\lambda} | \Phi_{\lambda}^{N\pm 1} \rangle$, where 
$| \Phi_{\lambda}^{N\pm 1} \rangle$ are given by Eq. (\ref{Np1}). This yields for the diagonal of 
the spectral function,
$A_{\lambda \lambda}(\omega)=2\pi\left[n_{\lambda} W^-_{\lambda}(\omega) +(1-n_{\lambda})W^+_{\lambda}(\omega) \right]$
with $W^{\pm}_{\lambda}(\omega)= \sum_j |\alpha^{\pm}_{j\lambda}|^2 \delta(\omega \pm [E_0^{N}-E_j^{N\pm 1}])$,
which has a finite width allowing to determine the quasi-particle life-times. Such investigation will be left to future.
Within this article we wish to focus on the, so called, density of states, which is obtained by taking the trace of 
the spectral function ($\sum_{\lambda} A_{\lambda\lambda}$):
\begin{eqnarray}\label{dos}
{\rm DOS}= 2\pi 
\sum_{\lambda}\left[ n_{\lambda} \delta(\omega-\epsilon^-_{\lambda})
+ (1-n_{\lambda}) \delta(\omega+\epsilon^+_{\lambda}) \right]
\end{eqnarray}

Now what remains is to calculate the excitation energies $\epsilon^{\pm}_{\lambda}=\epsilon^{\pm}_{i \bf k}=E_0^N - E_{i\bf k}^{N\pm1}$, where 
$E_{i \bf k}^{N \pm 1}$ is the energy of the system with an electron, with specific momentum {\bf k}, added/removed. 
While in experiments $E_{i\bf k}^{N \pm 1}$ represents the total energy of a macroscopic block of material, 
in the theoretical description $E_{i\bf k}^{N \pm 1}$ is the total
energy of a large but periodically repeated Born-von Karman (BvK) cell, where a constant charge background is added
to keep the total (infinite) system charge neutral. The calculation of such total energies is computationally 
very demanding; requiring the number of {\bf k}-points times the number of natural orbital (typically $\sim$2500)
constrained  ground-state calculations.  This is a formidable task and hence we make another simplification which 
is \emph{not conceptual} in nature 
but rather a numerical trick similar to the Slater transition state procedure \cite{slater}; we first introduce
the total ground-state energy, $E_{i\bf k}^{N \pm \eta}$, where a fractional number of particles, $\eta$, has been added/removed 
at a given {i\bf k}.
We then assume that upon adding/removing charge to {i\bf k} the only occupation number that changes significantly is the one 
that corresponds to the very same {i\bf k} while all the other occupation numbers as well as the natural orbitals remain unchanged. 
With this simplification, following Slater\cite{slater}, the $\epsilon^{\pm}$ can be approximated as 
\begin{align}\label{dedn}
 \epsilon^{\pm}_{i\bf k}= 
 \left.\frac{\partial E_{i\bf k}^{N \pm \eta}}{\partial \eta} \right|_{\eta=1/2}=
 \left.\frac{\partial E[\{\phi\},\{n\}]}
 {\partial n_{i\bf k}}\right|_{n_{i{\bf k}=1/2}}.
\end{align}
This simplification can be easily numerically validated by plotting the total energy as a function of $n_{i\bf k}$;
we find a nearly linear behavior for all the materials studied here. This implies that the 
Slater-type evaluation of the total-energy difference in Eq. (\ref{dedn}) is rather accurate.

Following the above procedure the spectral density for the strongly correlated Mott insulators
NiO, CoO, FeO and MnO is calculated using the full-potential linearized augmented plane wave (FP-LAPW) code Elk\cite{elk},
with practical details of the calculations following the scheme described in Ref.~(\onlinecite{sharma08}).

\begin{figure}[ht]
\centerline{\includegraphics[width=\columnwidth,angle=-0]{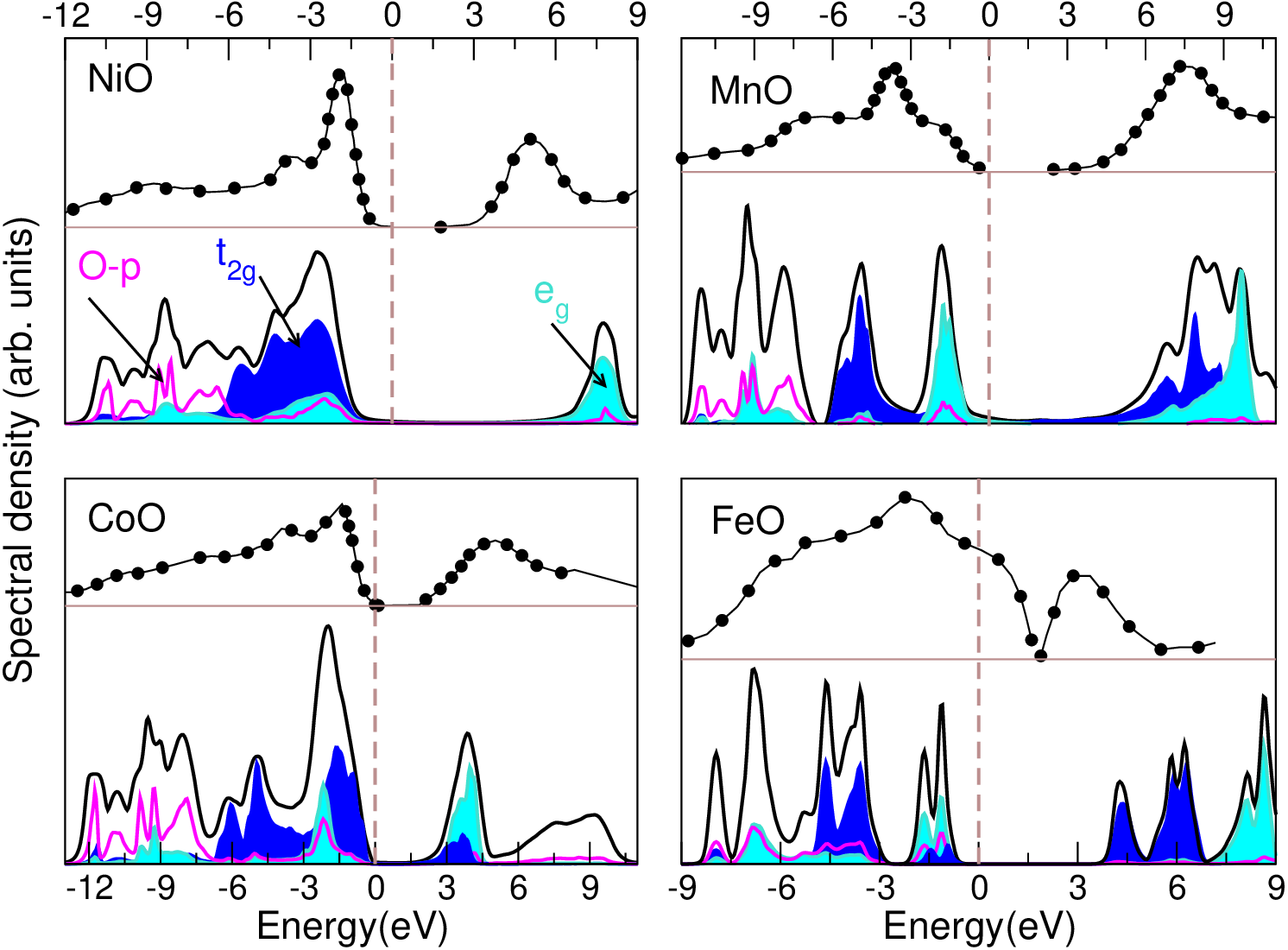}}
\caption{(Color online) Density of states for the TMOs in presence of AFM order.
Site and angular momentum projected spectral densities are also presented for transition metal $e_g$ and $t_{2g}$
states and Oxygen-$p$ states. In addition XPS and BIS spectra (shifted up for clarity) are
presented for comparison. Again, $\alpha=0.656$ for all materials. }
\label{tmo-mag}
\end{figure}

It is immediately apparent from Fig.~\ref{tmo-mag} that
RDMFT captures the essence of Mott-Hubbard physics: all the TMOs considered are insulating in nature. 
This fact was already noticed in the previous work \cite{sharma08} 
where the presence of a gap without any spin-order was deduced via a very 
different technique, namely the discontinuity in the chemical potential as a 
function of the particle number. A closer examination of the spectra for NiO, CoO and
MnO reveals an excellent agreement between the RDMFT peaks
and the corresponding XPS and BIS data. In fact, not only
the peak positions, but also their relative weights are well reproduced. 
For the case of FeO, it must be recalled that Fe segregation,
unavoidable in this compound, precludes the experimental realization
of pure FeO samples. For this reason the only existing experimental
data are rather old and the presumably substantially contaminated and broadened
data present no distinct features that may be used for a clear comparison.

The actual values of the insulating gaps that may be extracted from Fig.~\ref{tmo-mag}
are 4.5(4.3)eV, 2.6(2.8)eV, 3.2(3.6)eV, 3.5eV for NiO, CoO, MnO and FeO respectively with the 
corresponding experimental gap given in parenthesis. The value of the local moments we find to be 
1.36(1.9)$\mu_B$, 2.7(3.3)$\mu_B$, 3.35(3.62)$\mu_B$ and 3.38(4.7)$\mu_B$ for NiO, CoO, FeO and MnO 
respectively, again with the experimental values in parenthesis.
There are two reasons for the smaller values of the magnetic moment within RDMFT compared to experiment. 
Firstly, the calculations are performed with the FP-LAPW method 
in which space is divided into spheres around the atoms, the so called muffin-tins, and the interstitial region. 
In the case of fully non-collinear magnetic calculations the magnetic moment per site is calculated by integrating the 
magnetization vector field inside the muffin-tin. This means a
small part of the moment is lost in the interstitial. Secondly, the power functional induces a slight non-collinearity in the 
magnetization leading to yet more loss in the $z$-projected moment.

In Fig.~\ref{tmo-mag} we also present the site and angular momentum projected
spectral density for the TMOs considered in this work. The electronic gap, as expected,
always occurs between lower and upper Hubbard bands dominated by transition metal d-states. 
However, while for NiO one finds a significant component of oxygen-$p$ states in the 
lower Hubbard band, for the other TMOs this hybridization between oxygen-$p$ 
and TM-$d$ states reduces, and is almost absent in the case of
MnO, indicating that for this material the insulating state is driven
mostly by Mott-Hubbard correlations. 
As a validation of our method for the calculation of spectra we may compare these features of the 
projected spectral density, and in particular the ordering in energy of the $t_{2g}$ and $e_g$ states, 
with well established \emph{ab-initio} many-body techniques
such as DMFT and the $GW$ method\cite{kunes08,rodl}. In all cases we find excellent agreement, 
signaling that the method we present here yields an accurate description of the detailed features of the 
spectral density.

\begin{figure}[ht]
\vspace{0.5cm}
\centerline{\includegraphics[width=\columnwidth,angle=-0]{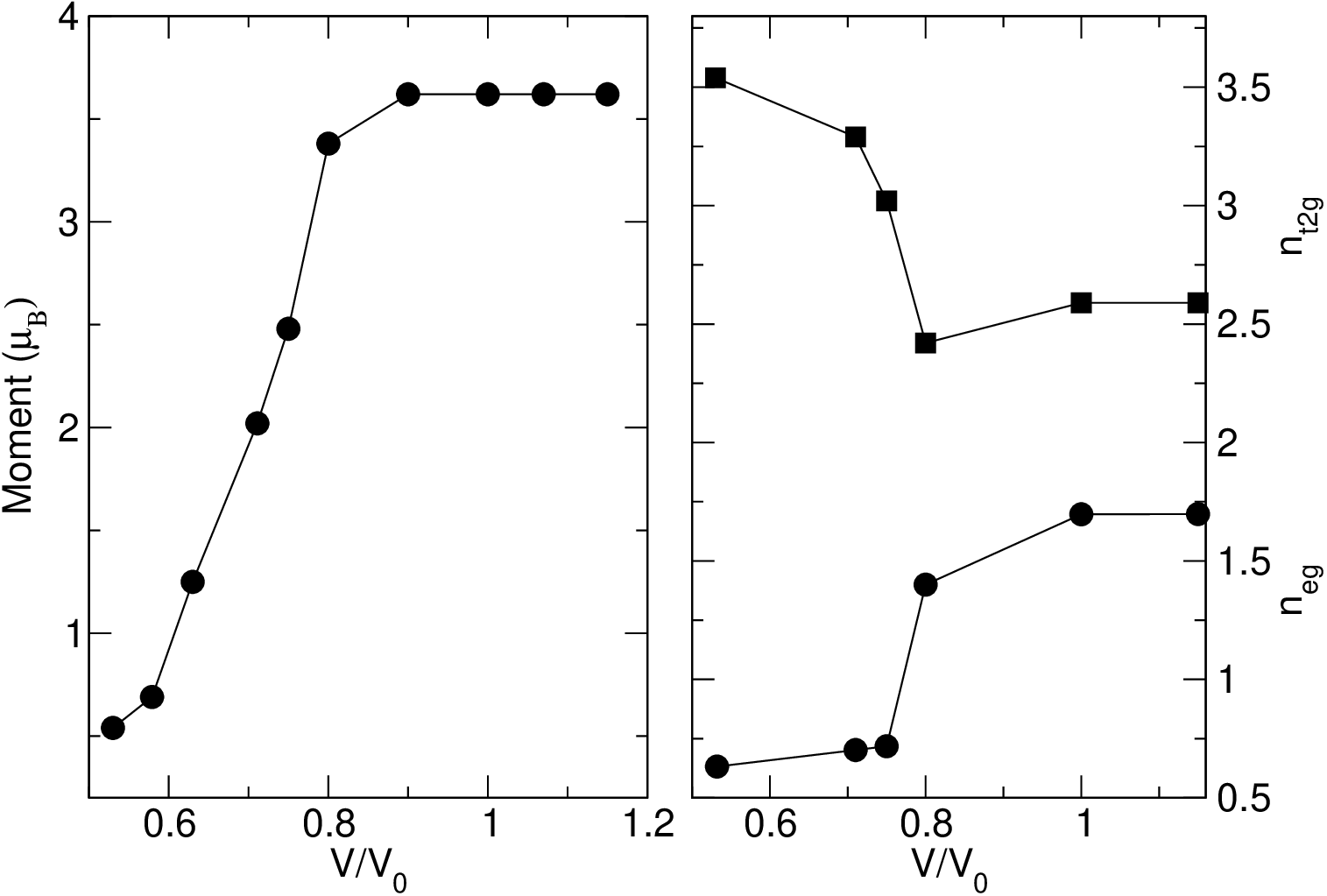}}
\caption{Left panel: on-site magnetic moment (in $\mu_B$) for MnO as a function of reduced volume ($v/v_0$). Right panel: 
Mn $d$-band occupancy resolved into $e_g$ and $t_{2g}$ components.}
\vspace{0.7cm}
\label{mno-mom}
\end{figure}

\begin{figure}[ht]
\vspace{0.5cm}
\centerline{\includegraphics[width=\columnwidth,angle=-0]{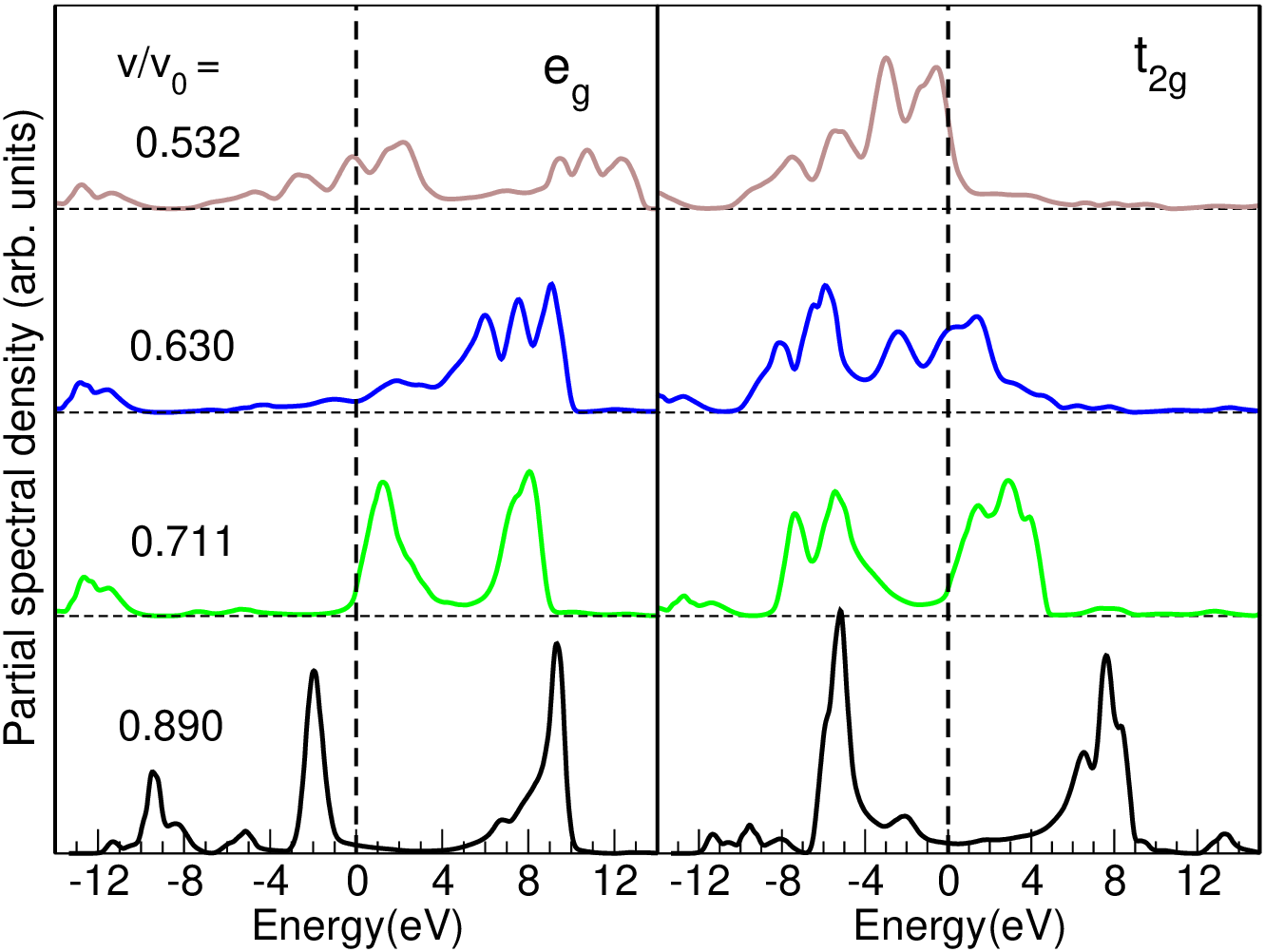}}
\caption{(Color online) $t_{2g}$ (right panel) and $e_g$ (left panel) projected spectral density for MnO as a function
of energy (in eV) and reduced volume. Results are obtained using RDMFT. Spectral density for different reduced volumes 
are shifted vertically for clarity.}
\vspace{0.7cm}
\label{mno-pres}
\end{figure}

Theoretical methods used to study the ground-state of TMOs agree with each other as far as the spectral density of 
TMOs at equilibrium volume is concerned\cite{rodl,luke,corbth,kunes08}, however, the actual values of the 
gap and the moment differ depending upon the details of the calculations. However, this agreement between 
various methods ends at ambient conditions, in that when attempting to study pressure induced
insulator-metal phase transitions (IMT) the results vary wildly depending upon the method. 
In MnO experimental data point towards a first order IMT which is accompanied 
by a reduced volume ($v/v_0$= 0.68 to 0.63) and moment collapse (5 to 1 $\mu_B$)\cite{expt1}. 
DFT with LDA/GGA like functionals captures this collapse of volume and magnetic moment and shows that the physics behind this 
phase transition is the simple widening of the Mn $d$-states due to increased itinerancy of the electrons at the reduced volume\cite{cohen,luke}.
On the other hand, DMFT shows that the phase transition occurs as a result of the increased crystal field splitting with the
width of the Mn $d$-states unchanged\cite{kunes08}. The picture obtained with correlated band theories is even more
complicated: LDA+$U$  shows a moment collapse but no IMT\cite{corbth}, hybrid functionals yield a phase transition to
a semi-metalic state\cite{corbth}, and finally the self-interaction corrected DFT shows a transition to a metallic state with
unusually large $d$-band width\cite{corbth}.

Given this wide spread of results, we study TMO phase transitions using RDMFT in the present work  to shed light on
this controversy. RDMFT is an ideal method for doing so as it is an \emph{ab-initio} theory for treating strongly correlated systems
not requiring any adjustable system dependent parameters.
The results, obtained using RDMFT, for the magnetic moment in MnO under applied pressure is shown in Fig. \ref{mno-mom}.
It is clear that the magnetic moment collapses from 3.6$\mu_B$ at optimal volume to 0.54$\mu_B$ at a reduced volume. 
Further reduction of the volume does not change the moment. 
Within RDMFT we find a volume collapse of 11\% which is higher than the experimental value of 6.6\%.
In order to investigate the prime reason behind this moment collapse we plot, in Fig. \ref{mno-mom}, the 
number of electrons in the Mn $d$-states as a function of volume. One notices a redistribution of electrons amongst the 
symmetry projected $t_{2g}$ and $e_g$ states. At the reduced volume of 0.8 the number of $e_g$  electrons ($n_{eg}$) starts
to reduce finally leveling off at v/v$_0$=0.711. This is accompanied by an increase in the $t_{2g}$ state charge ($n_{t2g}$). 
This picture is fully concomitant with the previous results obtained using DMFT\cite{kunes08}.
In order to look at the detailed behavior of these $t_{2g}$
and $e_g$ states we have also plotted them as a function of volume in Fig. \ref{mno-pres}. The $e_g$-states move above the chemical
potential as the volume is reduced, while the $t_{2g}$-states move below, finally ending up with a totally different spectral
density in the metallic phase as compared to the Mott insulating phase. As a result of this rearrangement the crystal field splitting 
between the $t_{2g}$ and $e_g$ states increases. Despite RDMFT being a totally different approach as compared to DMFT, the 
symmetry projected spectral density as a function of volume looks very similar for the two methods(see Fig. 3 of 
Ref. \onlinecite{kunes08}). However, there exists a
striking difference between the two in that within RDMFT the Mn $d$-states widen at reduced volume as compared to 
those at $v/v_0=1$.
This increase in the band width has its origin in a small shift in the spectral weight of the lower Hubbard band
to lower energy. A close inspection of Fig. \ref{mno-pres} shows that the change in crystal field splitting is very large
($\sim$4.4eV) and certainly has a prominent role in the IMT in MnO. Band widening is a co-existing phenomenon which has a very small 
influence on the IMT. This band widening on the other hand drives the IMT if LDA/GGA functionals with DFT are used. 

To conclude we have presented a method to calculate photo electron
spectra within the framework of RDMFT. We have shown that the spectral 
information obtained in this way gives a detailed account of the strongly correlated
nature of the TMOs, including the subtle interplay between Mott-Hubbard
correlation and charge-transfer character in these materials. We validate this method 
not only by showing a good agreement of gross spectral features with experiments, but also by 
a detailed comparison of the angular momentum projected spectral density
with that of well-established many-body techniques, in all cases finding an excellent agreement.
We have further elucidated the physics behind the insulator to metal phase transition in MnO using RDMFT. 
For MnO the pressure induced phase transition is caused by the increase in crystal field splitting which in turn
is the result of a re-distribution of charge amongst the states with $t_{2g}$ and $e_g$ symmetry. The widening of the 
transition metal $d$-band is  seen as a co-existing phenomenon, but is certainly not the reason for the
metalization.

\end{document}